\documentclass{article}
\usepackage[utf8]{inputenc}
\usepackage{url}
\usepackage{geometry}\providecommand{\keywords}[1]
{
  \small	
  \textbf{\textit{Keywords---}} #1
}

\usepackage[justification=centering]{caption}
\usepackage{amsmath}
\usepackage[table, dvipsnames]{xcolor}
\usepackage{amssymb}
\usepackage{xcolor}
\usepackage{mathptmx}       
\usepackage{helvet}         
\usepackage{courier}        
\usepackage{type1cm}        
\usepackage{makeidx}         
\usepackage{graphicx}        
\usepackage{tabularx}
\usepackage{float}
\usepackage{subcaption}
\usepackage{multirow}
\usepackage{textcomp}
\usepackage[linesnumbered,ruled]{algorithm2e}
\usepackage{multicol}        
\usepackage[bottom]{footmisc}

\title{DOC-NAD: A Hybrid Deep One-class Classifier for \\ Network Anomaly Detection}
\author{Mohanad Sarhan*$^{1}$, Gayan Kulatilleke$^{1}$, Wai Weng Lo$^{1}$, Siamak Layeghy$^{1}$,  Marius Portmann$^{1}$ \\
        \small $^{1}$University of Queensland, Brisbane, Australia \\
        \small *Corresponding Author: m.sarhan@uq.net.au\\
}

\date{}

\begin{document}

\maketitle

\begin{abstract}
Machine Learning (ML) approaches have been used to enhance the detection capabilities of Network Intrusion Detection Systems (NIDSs). Recent work has achieved near-perfect performance by following binary- and multi-class network anomaly detection tasks. Such systems depend on the availability of both (benign and malicious) network data classes during the training phase. However, attack data samples are often challenging to collect in most organisations due to security controls preventing the penetration of known malicious traffic to their networks. Therefore, this paper proposes a Deep One-Class (DOC) classifier for network intrusion detection by only training on benign network data samples. The novel one-class classification architecture consists of a histogram-based deep feed-forward classifier to extract useful network data features and use efficient outlier detection. The DOC classifier has been extensively evaluated using two benchmark NIDS datasets. The results demonstrate its superiority over current state-of-the-art one-class classifiers in terms of detection and false positive rates.
\end{abstract} \hspace{10pt}

\keywords{One-class classifier, intrusion detection, machine learning, anomaly detection}

\section{Introduction}
With the rapid increase and modification of network attack vectors, the need for dynamic attack detection is inevitable. Network attacks can cause information leakage, tampering, and disruption of organisational networks \cite{nazir2021survey}. Modern network attacks often require sophisticated tactics and techniques that may bypass current security control systems \cite{zhou2020unified}. Therefore, detecting network attacks is vital to maintaining the three principles of information security; confidentiality, integrity, and availability \cite{samonas2014cia}. Network Intrusion Detection Systems (NIDSs) aim to scan and detect network adversaries at the perimeter layer as they penetrate a computer network \cite{SARHAN2022}. There are two main types of NIDSs, i.e. signature- and anomaly-based NIDS \cite{jyothsna2011review}, that alert network and security administrators of a sign of a threat. Signature-based NIDSs scan incoming network traffic for any known Indicators of Compromise (IOCs), such as IPs, domain names, and hash values \cite{kumar2012signature}. This detection method is highly effective in identifying known threats as it compares traversing packets to a pre-configured list of known malicious signatures.

Anomaly-based NIDSs detect suspicious network behaviours as an anomaly by learning a computer network's normal or benign usage behaviour \cite{samrin2017review}. The baseline represents how the standard operating network traffic activities look, and any out-of-the-ordinary activity that does not fit the pre-defined profile will trigger a detection. Therefore, anomaly-based NIDSs are highly effective in detecting zero-day attacks compared to signature-based NIDSs \cite{sarhan2021zero} as they do not rely on pre-defined malicious signatures. Most modern anomaly-based NIDSs use Machine Learning (ML) techniques to train effective models for detecting network intrusions. ML-based NIDSs are designed with various architectures and are exposed to network data samples in the training phase to extract the semantic attributes of the captured training dataset. The training set is a collection of network data flows captured from the hosting organisation and presented in a flow export format such as NetFlow \cite{sarhan2020netflow}. During repeated training rounds, the models are optimised to minimise the error in mapping the input data samples to the desired output. 

There are two main and widely used types of ML classification; binary- and multi-class \cite{jha2019comparison}. Binary classification is widely adopted in developing ML-based NIDSs, with great success. The model learns the distinctive patterns between benign and malicious network traffic \cite{beaver2013learning}. The output predictions are in a binary format, indicating a safe or an unsafe test data sample. In the development of such models, the training dataset must include both data classes (malicious and benign) to facilitate a binary classification model \cite{yang2021bodmas}. Multi-classification has been used in developing ML-based NIDSs to classify network data traffic into benign, or one of the known attack classes \cite{singh2022survey}. The model learns the unique patterns of each available data class to identify the test data sample into one of them. Multi-class ML-based NIDSs are mainly used in network forensics to identify the type of exploit used in penetration. In developing multi-class models, the training dataset requires data samples from each data class.

The critical limitation of designing binary- and multi-class classification models is the dependence on the availability of each class data sample in the training phase \cite{abdulhammed2019features}. The necessity of having a training set consisting of both benign and malicious network data samples has made the development of such NIDSs challenging. Malicious data samples in production network environments are challenging to collect \cite{lohiya2020application} in sufficient amounts for ML training. This is due to most of the exploit attempts being blocked at the perimeter level of an organisation by a current security control such as a firewall or an existing NIDS. In case of a rare zero-day or successful exploit occurrence, accessing and capturing an adequate amount and quality of training data samples is difficult as the exact time of penetration and associated network data logs are unknown. Therefore, the requirement of collecting malicious network logs for ML training has significantly affected the development of ML-based NIDSs \cite{sarhan2023cyber}. One-class classification is an emerging technique of ML classification used in anomaly detection to overcome the limitations faced by other techniques, where the learning models are trained exclusively on data samples representing a single class \cite{khan2014one}.

This paper proposes a novel Deep One-Class (DOC) Classifier based on a one-class classification methodology for network anomaly detection. The DOC classifier solves the ever-growing challenge and limitation of other classification techniques of collecting sufficient attack traffic samples for training. The DOC classifier uses a deep one-class learning technique known as Deep Support Vector Data Description (Deep SVDD) \cite{pmlr-v80-ruff18a} to map network data features to an enhanced low-dimensional embedding, which is subsequently used by a Histogram-Based Outlier Score (HBOS) \cite{goldstein2012histogram} for anomaly detection. The DOC operates by building an accurate representative profile of the benign network standard operating activities and detects outliers as an anomaly. Powerful deep feature extraction and rapid HBOS detection enable the DOC classifier to detect network intrusions effectively and accurately. DOC development requires benign network data samples only for the training process and omits the requirement of attack data collection, which makes it a practically suitable ML-based NIDS. 

The paper is structured as follows; In Section \ref{rw}, key related works that use one-class classification techniques to develop ML-based NIDSs are discussed. In Section \ref{main}, the usage of the proposed DOC classifier is motivated, and its architecture is explained. The evaluation methodology of the DOC classifier on two widely used NIDSs is described in Section \ref{ev}. The results of the DOC detection performance are presented in Section \ref{r} and compared with five key one-class classifiers. The main contributions of this paper are a) the design of a novel DOC classifier for NIDS that does not require the availability of attack data samples for training and b) the extensive evaluation of the DOC classifier across two datasets and comparison with other classifiers demonstrating the superiority of the proposed framework.

\section{Related Work}
\label{rw}
This section reviews some of the key related work proposing one-class classifiers for NIDS. Kind et al. \cite{kind2009histogram} highlighted the promising uses of histograms in the detection of network intrusions. The method proposed in 2009 involves constructing histograms of different network traffic features. Each histogram is embedded into a metric space to position similar histograms together. Data mining techniques are used to model benign network behaviour. The typical patterns are compared with the incoming network traffic feed to identify deviations and detect anomalies. Two real-world and one synthetic datasets were used to evaluate the proposed framework. The results demonstrate the superiority of histograms over entropy-based distribution approximations in detecting a wide range of anomalies.

Zavrak et al. \cite{zavrak2020anomaly} proposed a variational autoencoder system for network anomaly detection. The authors use flow-based features because of their ease of extraction for model training and compare the detection performance with traditional autoencoder and one-class Support Vector Machine (SVM) models. In a one-class classification methodology, models were trained on benign-only data samples obtained from the CICIDS2017 NIDS dataset. It is shown that the variational autoencoder-based anomaly detection system performs better compared to the other considered models. The proposed method achieved a higher detection rate in 9 of the 14 attack groups available in the dataset.

An evaluation study \cite{arregoces2022network} compares the performance of several one-class classifier models using the UNSW-NB15 dataset. The authors highlight the necessity of adopting the one-class classification methodology in developing NIDSs due to the imbalanced nature of the datasets. The evaluation involved developing and evaluating one-class SVM, isolation forest, minimum covariance determinant, and local outlier factor models. The Pearson correlation feature selection technique discarded the highly correlated data features and used the random forest to identify the most relevant ones. The results demonstrate the superiority of one-class SVM over the other considered approaches with an accuracy of 61.9\%.

In \cite{verkerken2020unsupervised}, Verkerken et al. evaluated the performance of various unsupervised ML models using the CIC-IDS-2017 NIDS dataset. The paper highlights the importance of unsupervised techniques in detecting zero-day attack groups. The dataset is transformed using Principal Component Analysis (PCA) for dimensionality reduction. In a one-class classification technique, the models were trained on benign-only data samples and evaluated on merged (benign and malicious) data samples. The results show that autoencoders yield the best detection performance of a 96.16\% F1 score, followed by one-class SVM, isolation forest, and PCA classifiers. 

Zhang et al. \cite{zhang2015anomaly} recommend using one-class SVM over the traditional two-class variant due to the ease of constructing the training sets in developing NIDSs. Both classifiers were trained and tested using the KDDCUP99 dataset and compared to the performance of a Probabilistic Neural Network (PNN). The one-class SVM trained on the benign data samples achieved an effective detection rate on the DoS and Probe attack categories; however, the R2L, U2R, and "other" attack groups were unreliably detected. In comparison, the one-class SVM achieved a higher overall detection rate but a lower precision rate due to a more significant number of false positives compared to the PNN and two-class SVM.

In \cite{vasudevan2016local}, Vasudevan et al. combined the advantages of supervised and unsupervised learning to obtain better intrusion detection performance. The proposed hierarchical model is devised in three stages: Dirichlet Process (DP) clustering based on the underlying data distribution, Local Outlier Factor (LOF) to identify dense local areas and group them into four bins, and finally, a one-class classifier to model the benign instances in each bin. Hybrid density estimation, reconstruction, and boundary methods are used in the design of the one-class classifier. The evaluation of the proposed model is carried out on the KDDCup99 and SSENet-2011 datasets and compared with one-class SVM. The results demonstrate the superiority of the proposed model on both datasets.

\section{Hybrid DOC}
\label{main}

Current ML-based NIDSs implement ML models in binary- and multi-class classification techniques, which require a training set constructed of benign and malicious data samples. This practice has been deemed challenging in the real world, as the practical collection of sufficient anomaly data samples in a traffic trace is complex. This is because most known attacks are generally blocked at the perimeter by current security controls. In the rare occasion of a successful exploit, the complete set of associated network traffic is unknown, and their real-time capture is difficult. The collection of zero-day attacks for ML training is impossible due to their non-existence at the time of ML-based NIDS development. Furthermore, training and successful evaluation of ML models on known attacks does not indicate a generalisable behaviour on unknown attacks, as illustrated in \cite{sarhan2021zero}. Therefore, the practical deployments of effective ML-based NIDSs in binary- and multi-class classification techniques are limited.

One-class classification is an emerging paradigm of ML and is generally associated with anomaly or outlier detection. The models are exclusively trained on "normal" class data samples to build a profile of the accepted usage. One-class classification assumes that the training set is free or insufficiently sampled of the "abnormal" class data samples. This assumption correlates with the real-world networking application, as the benign (normal) class data samples are easy to capture and analyse in an operational network environment. Whereas, on the other hand, it is very challenging to collect a sufficient amount of malicious (abnormal) class data samples. More importantly, due to the rapid increase and modification of network attack vectors, it is impossible to collect malicious network data samples representing the wide range of attack scenarios, such as zero days, that will be observed in production (post-training). Therefore, from practical experience, it is easier to train a one-class classifier compared to binary- and multi-class classifiers in ML-based NIDS models. 

\begin{figure}[!h]
  \centering
  \includegraphics[width=0.85\textwidth]{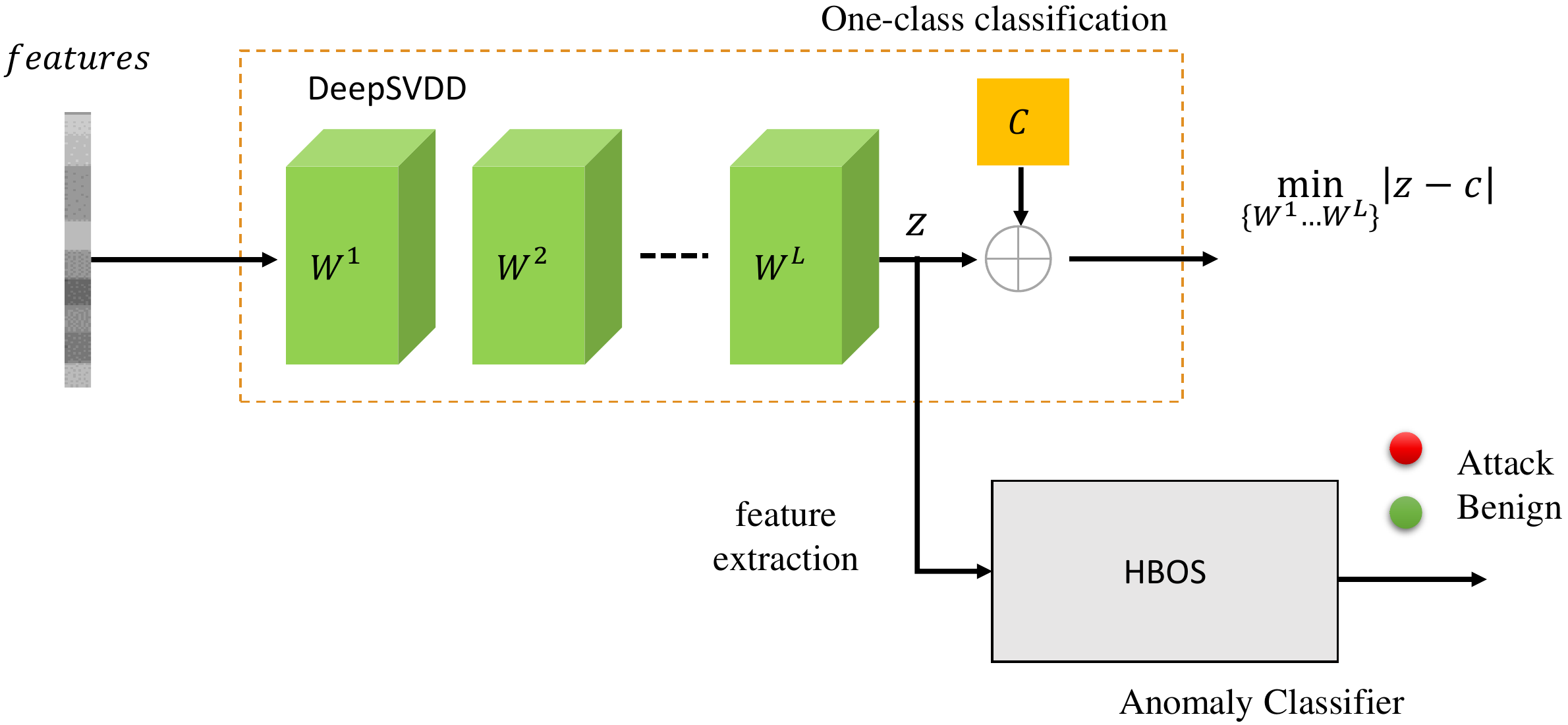}  
  \caption{System architecture}
  \label{arc}
\end{figure}
The DOC classifier proposed in this paper is designed using a hybrid one-class architecture that utilises the deep extraction of data point embeddings and the construction of histograms of benign network data samples. The training stage involves mapping extracted semantic attributes to an accurate representative profile of the standard operating network environment. Outlier network flows that do not fit within the set profile are detected as anomalies. The DOC architecture consists of two main integrated components: DeepSVDD \cite{pmlr-v80-ruff18a}, and HBOS \cite{goldstein2012histogram}, as shown in Figure \ref{arc}. The integration of the two components complements the benign data deep extraction and mapping process to improve the anomaly detection performance compared to other classifiers.

The first component of DOC is known as DeepSVDD, which uses deep learning to extract relevant features automatically without the substantial feature engineering processes required in shallow learning methods. DeepSVDD is inspired by a kernel-based one-class classification, and minimum volume estimation \cite{pmlr-v80-ruff18a}. During training, the neural model minimises the hypersphere volume that encloses the network representations of the benign data samples in the output space. This technique forces the model to extract the common factors of variation of the data distribution as the data points are closely mapped to the centre of the sphere. Stochastic Gradient Descent (SGD) is used to optimise the parameters of the neural network model using back-propagation.

The main objective of DeepSVDD is defined in Equation \ref{eq2}, where the input and output spaces are $\mathcal{X} \subseteq \mathbb{R}^d$ and  $\mathcal{F} \subseteq \mathbb{R}^p$ respectively, $\phi(\cdot ; \mathcal{W}): \mathcal{X} \rightarrow \mathcal{F}$ is a neural network containing $L \in \mathbb{N}$ hidden layers and set of weights $\mathcal{W}=\left\{\boldsymbol{W}^1, \ldots, \boldsymbol{W}^L\right\}$. Specifically, $\phi(\boldsymbol{x} ; \mathcal{W}) \in \mathcal{F}$ is the learnt representation of $\boldsymbol{x} \in \mathcal{X}$ from network $\phi$ with parameters $\mathcal{W}$. DeepSVDD aims to learn $\mathcal{W}$ by finding a hypersphere of minimum volume with centre $c \in \mathcal{F}$. The hypersphere is contracted by minimising the mean distance of all data representations to the centre $c$. Essentially, minimizing $\left\|\phi\left(\boldsymbol{x} ; \mathcal{W}^*\right)-\boldsymbol{c}\right\|^2$ minimizes the volume of the hypersphere. The second term is a weight decay regulariser with hyperparameter $\lambda>0$. $\|\cdot\|_F$ denotes the Frobenius norm. 

\begin{equation}
\min _{\mathcal{W}} \frac{1}{n} \sum_{i=1}^n\left\|\phi\left(\boldsymbol{x}_i ; \mathcal{W}\right)-\boldsymbol{c}\right\|^2+\frac{\lambda}{2} \sum_{\ell=1}^L\left\|\boldsymbol{W}^{\ell}\right\|_F^2
\quad \operatorname{where} \mathcal{W} = \{W^1 \cdots W^L\}.
\label{eq2}
\end{equation}

Optimising Equation \ref{eq2} results in benign network data points being closely mapped to the centre $c$ of the hypersphere through learning to extract the common factors of variation of the data. While benign data samples will be closely mapped to $c$, outliers (which lack the common factors) will be mapped further away. DeepSVDD assumes most of the training data $\mathcal{D}_n$ is normal, which is a reasonable assumption in one-class classification tasks and, it penalises the mean distance over all data points forcing the benign data points closer to $c$. We use the representations $z_i$ from the trained DeepSVDD obtained via Equation \ref{eq1}.

\begin{equation}
z_i = \phi\left(\boldsymbol{x}_i ; \mathcal{W}\right)
\label{eq1}
\end{equation}

The second component of DOC, known as HBOS, maps the features extracted by DeepSVDD into a set of histograms. A histogram is a statistical distribution of several samples over the possible values of the data value. In a NIDS dataset, histograms can present the distribution of a network traffic data feature such as the number of TCP flags, the TTL value or in/out bytes. Histograms model the detailed data features of network traffic, which enables the identification of a broader range of anomalies. HBOS assumes independence of data features and calculates the degree of anomaly by building histograms. This feature makes HBOS computationally efficient compared to multivariate approaches, which makes it suitable for developing NIDS where the detection results are required immediately.

HBOS constructs a univariate histogram for each network data feature. A static number of bins are used using $k$ equal width bins over the value range. The frequency of samples located in each bin is used to estimate the height of the bins. As shown in Equation \ref{eq3}, for each data feature $d$, a histogram is constructed with a height representing the density estimation. All histograms are normalised to a maximum height of 1.0, ensuring an equal weight of each feature to the anomaly score. Finally, the HBOS of each data sample $p$ is calculated using the corresponding height of the bins where the sample is located.

\begin{equation}
H B O S( z_i )=\sum_{j=0}^d \log \left(\frac{1}{\text { hist }_j(z_i)}\right)
\label{eq3}
\end{equation}

\begin{figure}[!h]
  \centering
\includegraphics[width=0.85\textwidth]{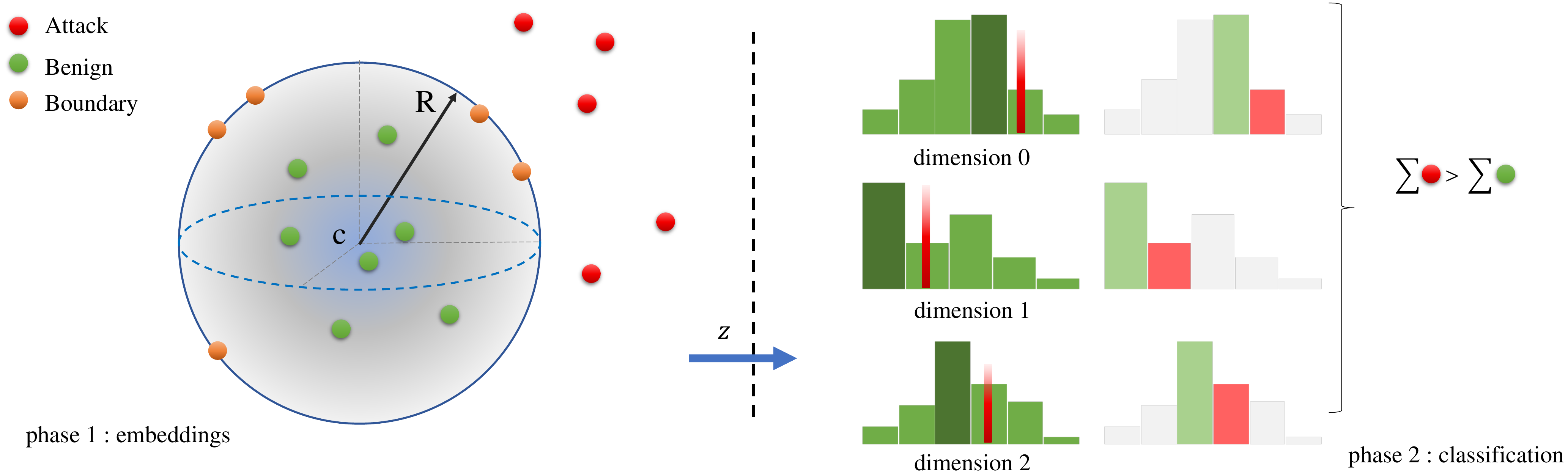}  
  \caption{The two-phase Benign vs Anomaly data point representations and classification}
  \label{badp}
\end{figure}

In Figure~\ref{badp}, the two-phase data point representation generation and classification of the DOC are illustrated. The left side displays the first phase where the deep one-class classification objective of DeepSVDD (Equation~\ref{eq2}) is used to generate discriminative lower $d$ dimensional feature representations. Essentially, deep extraction is performed by learning a mapping function $\phi$ such that the majority benign data samples are closer towards the centre $c$ of the enclosing hyper-sphere of a radius $R$ while the  minority anomaly data samples tend to be further away.
In the next phase, shown on the right side, learnt representations $z_i$ are used to build a probability distribution for \textit{each} lower dimensional feature $j \in \{1,\dots,d\}$ of the majority benign representations. This expected set of distributions is shown in green.

Classification is performed via the HBOS score, which is obtained by aggregating each of the feature's inverse occurrence probability obtained from the expected distribution of the majority benign distribution. In a benign sample, the feature values are closer to the expected probability distribution, and the aggregated inverse HBOS score is lower. However, as anomalous samples tend to differ from the majority distribution (as shown in red), and as a result, probabilities of feature values are lower and the aggregated inverse HBOS score is higher.

\section{Evaluation Methodology}
\label{ev}

The DOC model is evaluated on two modern NIDS datasets. Several other one-class classifiers are implemented and evaluated to compare the detection performance on the same datasets. The most common and widely used models are implemented for comparison. The chosen models are Isolation Forest (IF) \cite{liu2008isolation}, Principal Components Analysis (PCA) \cite{shyu2003novel}, and Variational Autoencoder (VAE) \cite{kingma2013auto}. In addition, we utilise the standalone DeepSVDD and HBOS models as a benchmark to the developed DOC model. The experiments were run in Python using PYOD \cite{zhao2019pyod} based on Keras for ML and Pandas for data processing. The default hyperparameters of the PYOD library were used in designing the models. Standard NIDS classification performance metrics are used to evaluate the one-class classifiers, i.e. Detection Rate (DR), False Alarm Rate (FAR), Area Under the Curve (AUC), and F1 score. These metrics are defined based on the numbers of True Positives (TP), False Positives (FP), True Negatives (TN) and False Negatives (FN), as shown in Table \ref{eme}.

\renewcommand{\arraystretch}{1.5}
\begin{table}[!h]\scriptsize	
\centering
\caption{Evaluation Metrics}
\begin{tabular}{|>{\centering\arraybackslash}m{4cm}|>{\centering\arraybackslash}m{5.5cm} |>{\centering\arraybackslash}m{3cm} |}

\hline
\textbf{Metric}            & \textbf{Definition}           & \textbf{Equation}                               \\ \hline
Accuracy                   & The percentage of correctly classified samples in the test set.              &\normalsize $\frac{TP+TN}{TP+FP+TN+FN} \times 100$\\ \hline
Detection Rate (DR)        & The percentage of correctly classified total attack samples in the test set. &\normalsize $\frac{TP}{TP+FN} \times 100 $\\ \hline

False Alarm Rate (FAR)     & The percentage of incorrectly classified benign samples in the test set. &  \normalsize $\frac{FP}{FP+TN} \times 100$   \\ \hline

Area Under the Curve (AUC) & The area underneath the DR and FAR plot curve in the test set.        & N/A       \\ \hline
F1 Score                   & The harmonic mean of the model's precision and DR.  & \normalsize $2 \times \frac{DR\;\times \;Precision}{DR\; +\; Precision}$         \\ \hline 
\end{tabular}%
\label{eme}
\end{table}
\renewcommand{\arraystretch}{1}

Two key and widely used NIDS datasets are used to evaluate the one-class classifiers, i.e., NF-UNSW-NB15-v2  and NF-CSE-CIC-IDS2018-v2 \cite{sarhan2022towards}. Both datasets are synthetic and created via virtual network testbeds representing organisational network environments. Synthetic datasets are widely used in the literature as they overcome the privacy and security concerns encountered in real-world production networks. Benign network traffic is captured for standard network usage baseline. Moreover, several attack scenarios are conducted, and the corresponding network traffic is collected. Network traffic is captured in its native packet capture (pcap) format. Further processing involves the extraction of network data features for traffic analysis. The features present information regarding the data flows, labelled malicious or benign records. The network data flows in NF-UNSW-NB15-v2, and NF-CSE-CIC-IDS2018-v2 are presented in NetFlow v9 standard format. NetFlow is a de facto flow network monitoring and analysis standard due to its practicality and ubiquitous deployment. Each used dataset has been generated over different test beds and includes a unique set of benign applications and malicious use cases. This consequently results in a variation of the statistical distributions held by each dataset. Therefore, the datasets used in this paper are non-Independent and Identically Distributed (IID) and extensively evaluate the proposed model.

\begin{itemize}
    \item \textbf{NF-UNSW-NB15-v2 \cite{sarhan2022towards}}- A NetFlow dataset based on the UNSW-NB15 dataset has recently been generated and released in 2021. The original dataset was released in 2015 by the Cyber Range Lab of the Australian Centre for Cyber Security (ACCS). The IXIA Perfect Storm tool was configured to simulate benign network traffic and synthetic attack scenarios. The dataset is generated by extracting 43 NetFlow-based features, explained in \cite{sarhan2022towards}, from the pcap files of the UNSW-NB15 dataset. The nprobe feature extraction tool extracts network data flows, which are labelled using the appropriate data labels. The total number of data flows is 2,390,275, of which 95,053 (3.98\%) are attack samples and 2,295,222 (96.02\%) benign. There are nine attack groups: Exploits, Fuzzers, Generic, Reconnaissance, DoS, Analysis, Backdoor, Shellcode, and Worms.
    
    \item \textbf{NF-CSE-CIC-IDS2018-v2 \cite{sarhan2022towards}}- An IoT NetFlow-based dataset released in 2021 containing different attack types, such as brute-force, bot, DoS, DDoS, infiltration, and web attacks. The exploits are conducted from an external network to simulate realistic attack scenarios. The dataset is generated by converting the publicly available pcap files of the CSE-CIC-IDS2018 \cite{sharafaldin-habibi-lashkari-ghorbani-2018s} dataset to 43 NetFlow v9 features using the nprobe \cite{deri2003nprobe} tool. The total number of data flows is 18,893,708, of which 2,258,141 (11.95\%) are attack samples and 16,635,567 (88.05\%) benign ones. The source dataset (CSE-CIC-IDS2018) was released by a collaborative project between the Communications Security Establishment (CSE) and the Canadian Institute for Cybersecurity (CIC) in 2018. Their developed tool called CICFlowMeter-V3 was used to extract 75 network data features. The network testbed simulates a realistic organisational computer network consisting of five departments and a server room.
\end{itemize}

The publicly available datasets are further processed in this paper for efficient ML operation and reliable evaluation. Initially, the flow identifiers, such as source/destination IPs and ports, are removed to avoid learning bias towards the attacking- and victim-end nodes. This is due to the nature of synthetic network datasets where distinct nodes are used in launching malicious traffic. The benign samples are split into training and testing sets in a ratio of 70\% to 30\%, respectively. The attack samples are only added to the testing sets to accommodate for one-class classification methods. Finally, the Min-Max Scaler normalisation technique is applied to obtain all values between 0 and 1 for effecting ML training. The normalisation occurs by applying  \begin{equation}X_*=\frac{X-X_{min}}{X_{max}-X_{min}}\end{equation}
\noindent on each set, where $X\textsubscript{*}$ represents the final output value ranging from 0 to 1. $X$ is the original input value, and $X\textsubscript{max}$ and $X\textsubscript{min}$ indicate the maximum and minimum values for each feature, respectively. To obtain reliable and fair evaluation metrics, a k-fold cross-evaluation technique is adopted with $k=5$.

\section{Results}
\label{r}
The proposed DOC model is implemented and evaluated across two NIDS datasets, and the detection performance is compared with five one-class classifiers. In Table \ref{tab:un}, the evaluation metrics achieved by the classifiers on the NF-UNSW-NB15-v2 dataset are presented. The accuracy and F1 score metrics of IF, PCA, and VAE are similar; they all achieved around 90\%, caused by a low DR and a high FAR. The DeepSVDD and HBOS achieved similar performance with an increased DR compared to the other classifiers. However, the FAR remained high at around 10\%, increasing the AUC to ~95\%. The proposed DOC classifier has achieved a significantly lower FAR than the rest of the classifiers at a rate of 2.01\% while maintaining a high DR. The DOC classifier with an F1 score and AUC of 98.26\% and 98.89\%, respectively, achieved the best evaluation metrics.

\begin{table}[h!]\footnotesize
\centering
\caption{NF-UNSW-NB15-v2 results}
\label{tab:un}
\begin{tabular}{l|r|r|r|r|r|}
\cline{2-6}
\textbf{}                               & \multicolumn{1}{l|}{\textbf{Accuracy}} & \multicolumn{1}{l|}{\textbf{F1 Score}} & \multicolumn{1}{l|}{\textbf{AUC}} & \multicolumn{1}{l|}{\textbf{DR}} & \multicolumn{1}{l|}{\textbf{FAR}} \\ \hline
\multicolumn{1}{|l|}{\textbf{IF}}       & 89.87                                  & 90.83                                  & 89.45                             & 88.90                             & 10.00                              \\ \hline
\multicolumn{1}{|l|}{\textbf{PCA}}      & 88.85                                  & 89.81                                  & 85.22                             & 80.42                            & 9.98                              \\ \hline
\multicolumn{1}{|l|}{\textbf{VAE}}      & 88.91                                  & 89.87                                  & 85.54                             & 81.09                            & 10.01                             \\ \hline
\multicolumn{1}{|l|}{\textbf{DeepSVDD}} & 91.26                                  & 92.19                                  & 95.03                             & 100.00                            & 9.94                              \\ \hline
\multicolumn{1}{|l|}{\textbf{HBOS}}     & 91.23                                  & 92.16                                  & 95.01                             & 100.00                            & 9.98                              \\ \hline
\multicolumn{1}{|l|}{\textbf{DOC}}      & \textbf{98.20}                         & \textbf{98.26}                         & \textbf{98.89}                    & \textbf{99.79}                   & \textbf{2.01}                     \\ \hline
\end{tabular}%
\end{table}

Table \ref{tab:cic} presents the results of the one-class classifiers achieved on the NF-CSE-CIC-IDS2018-v2 dataset. IF and VAE performance is the poorest across the range of classifiers, caused by low attacks DR with an AUC of 74.90\% and 76.87\%, respectively. The rest of the classifiers (PCA, DeepSVDD, HBOS, and DOC) achieve similar detection performance, with DOC leading the range with an AUC of 85.52\%. The proposed DOC classifier achieves the best performance across all metrics except for the DR, with PCA having a higher attack detection value of 81.60\% compared to 76.03\%. However, a significantly lower FAR of 5.00\% is noticed by the DOC classifier, which makes it a superior classifier overall.

\begin{table}[h!]\footnotesize
\centering
\caption{NF-CSE-CIC-IDS2018-v2 results}
\label{tab:cic}
\begin{tabular}{l|r|r|r|r|r|}
\cline{2-6}
\textbf{}                               & \multicolumn{1}{l|}{\textbf{Accuracy}} & \multicolumn{1}{l|}{\textbf{F1 Score}} & \multicolumn{1}{l|}{\textbf{AUC}} & \multicolumn{1}{l|}{\textbf{DR}} & \multicolumn{1}{l|}{\textbf{FAR}} \\ \hline
\multicolumn{1}{|l|}{\textbf{IF}}       & 80.60                                  & 80.01                                  & 74.90                             & 59.80                             & 9.99                              \\ \hline
\multicolumn{1}{|l|}{\textbf{PCA}}  &    87.39                                  & 87.45                                  & 85.41                             & 81.60                            & 9.99                             \\ \hline
\multicolumn{1}{|l|}{\textbf{VAE}}      & 81.83                                  & 81.42                                  & 76.87                             & 63.72                           & 9.98                             \\ \hline
\multicolumn{1}{|l|}{\textbf{DeepSVDD}} & 86.01                                 & 86.00                                 & 83.59                          & 77.16                            & 9.98                             \\ \hline
\multicolumn{1}{|l|}{\textbf{HBOS}}     & 85.61                                  & 85.57                                  & 82.91                             & 75.72                            & 9.91                             \\ \hline
\multicolumn{1}{|l|}{\textbf{DOC}}      & \textbf{89.09}                         & \textbf{88.87}                         & \textbf{85.52}                    & \textbf{76.03}                   & \textbf{5.00}                     \\ \hline
\end{tabular}%
\end{table}

The proposed DOC classifier achieves the best-performing detection metrics across two different NIDS datasets compared to the rest of the considered classifiers. A great feature of the DOC classifier is the low FAR value, where a significant decrease of around 80\% and 50\% were observed on the NF-UNSW-NB15-v2 and NF-CSE-CIC-IDS2018-v2 datasets, respectively. This is an essential improvement considering how a high number of FP alerts would result in security operation teams' fatigue and, eventually lack of trust in the triggered events. While attaining a low FAR, DOC maintained a competitive DR equal to or slightly lower than the rest. Overall, the superior performance of the DOC classifier across the different ranges of classification metrics confirms the efficiency and effectiveness of the proposed architecture.

\section{Conclusion}
In this paper, a deep one-class classification model is proposed and defined as DOC. The model is evaluated using two key NIDS datasets, and its performance is compared to five standard one-class classifiers. The proposed approach reliably detects unseen attack groups and outperforms the state-of-the-art classifiers. A high level of DRs and significantly lower FARs across the two datasets demonstrate the effectiveness of the DeepSVDD and HBOS integration. The deep extraction and mapping of the benign network environment to univariate histograms is a promising feature of ML-based NIDSs and an effective solution to the lack of malicious data samples challenge and should be further explored and evaluated.


\bibliography{main.bib}

\end{document}